\documentclass[conference]{IEEEtran}
\IEEEoverridecommandlockouts
\usepackage{cite}
\usepackage{amsmath,amssymb,amsfonts}
\usepackage{algorithmic}
\usepackage{graphicx}
\usepackage{textcomp}
\usepackage{xcolor}
\usepackage{arydshln}
\usepackage{url}

\def\BibTeX{{\rm B\kern-.05em{\sc i\kern-.025em b}\kern-.08em
    T\kern-.1667em\lower.7ex\hbox{E}\kern-.125emX}}

\makeatletter
\newcommand{\linebreakand}{%
  \end{@IEEEauthorhalign}
  \hfill\mbox{}\par
  \mbox{}\hfill\begin{@IEEEauthorhalign}
}
\makeatother
    
\begin{document}

\title{Signed Ego Network Model and its Application to Twitter
\thanks{\fbox{
\parbox{\dimexpr\columnwidth-20pt}{
\copyright \textbf{978-1-6654-8045-1/22/\$31.00 ©2022 IEEE}. Personal use of this material is permitted.  Permission from IEEE must be obtained for all other uses, in any current or future media, including reprinting/republishing this material for advertising or promotional purposes, creating new collective works, for resale or redistribution to servers or lists, or reuse of any copyrighted component of this work in other works.}}
}}

\author{\IEEEauthorblockN{Jack Tacchi}
\IEEEauthorblockA{\textit{Istituto di Informatica e Telematica} \\
\textit{Consiglio Nazionale delle Richerche}\\
Pisa, Italy \\
jack.tacchi@sns.it}
\and
\IEEEauthorblockN{Chiara Boldrini}
\IEEEauthorblockA{\textit{Istituto di Informatica e Telematica} \\
\textit{Consiglio Nazionale delle Richerche}\\
Pisa, Italy \\
chiara.boldrini@iit.cnr.it}
\linebreakand
\IEEEauthorblockN{Andrea Passarella}
\IEEEauthorblockA{\textit{Istituto di Informatica e Telematica} \\
\textit{Consiglio Nazionale delle Richerche}\\
Pisa, Italy \\
andrea.passarella@iit.cnr.it}
\and
\IEEEauthorblockN{Marco Conti}
\IEEEauthorblockA{\textit{Istituto di Informatica e Telematica} \\
\textit{Consiglio Nazionale delle Richerche}\\
Pisa, Italy \\
marco.conti@iit.cnr.it}
}

\maketitle

\begin{abstract}
The Ego Network Model (ENM) describes how individuals organise their social relations in concentric circles (typically five) of decreasing intimacy, and it has been found almost ubiquitously in social networks, both offline and online. The ENM gauges the tie strength between peers in terms of interaction frequency, which is easy to measure and provides a good proxy for the time spent nurturing the relationship. However, advances in signed network analysis have shown that positive and negative relations play very different roles in network dynamics. For this reason, this work sets out to investigate the ENM when including signed relations. The main contributions of this paper are twofold: firstly, a novel method of signing relationships between individuals using sentiment analysis and, secondly, an investigation of the properties of Signed Ego Networks (Ego Networks with signed connections). Signed Ego Networks are then extracted for the users of eight different Twitter datasets composed of both specialised users (e.g. journalists) and generic users. We find that negative links are over-represented in the active part of the Ego Networks of all types of users, suggesting that Twitter users tend to engage regularly with negative connections. Further, we observe that negative relationships are overwhelmingly predominant in the Ego Network circles of specialised users, hinting at very polarised online interactions for this category of users. In addition, negative relationships are found disproportionately more at the more intimate levels of the ENM for journalists, while their percentages are stable across the circles of the other Twitter users.
\end{abstract}


\section{Introduction}
\label{sec:intro}
Humans are social animals. Every day we interact with other people and these interactions link us to each other to an extent that is not immediately obvious without observing the social network as a whole. Since the advent of the internet, people have been able to interact with each other with far greater ease and regardless of geographical location. In recent years and with the rise of online social networks (OSNs), communications have been facilitated even further. Indeed, it is now possible to instantly interact with someone on the other side of the globe with the mere click of a button. This increasingly global connectivity has made it more important than ever to be able to understand social networks and the interactions that take place in them.

\begin{figure}
    \centerline{\hspace{50pt}\includegraphics[scale=0.22]{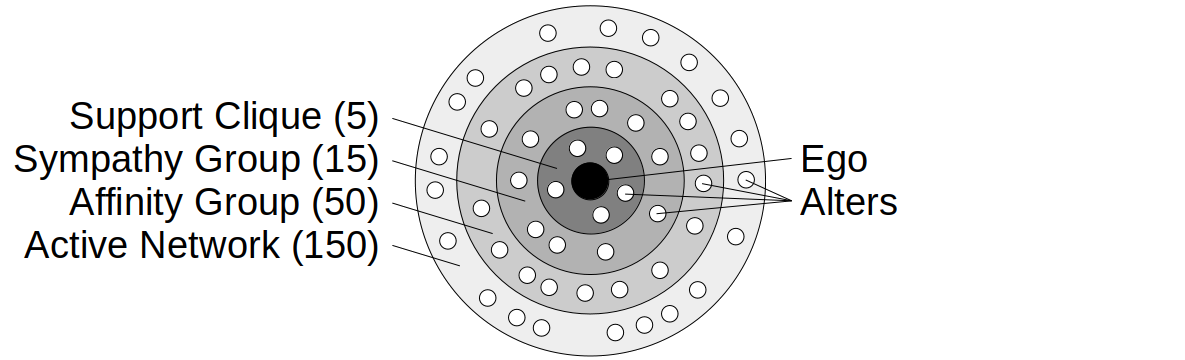}}
    \caption{The Ego Network Model, with the names and expected sizes of each subgroup for social networks of humans.}
    \label{Ego_Network_Model}
\end{figure}

Significant effort has been devoted to modelling the structural properties of relationships in social networks, with a significant share using graph-based models. One such representation is the Ego Network Model (ENM), based on findings from evolutionary anthropology about how humans organise their social relationships~\cite{Dunbar_1995}. The ENM model takes the perspective of a single user, the \emph{Ego}, who is placed at the centre of the model and surrounded by all of their immediate connections, named \emph{Alters}, who are organised based on their tie strength to the Ego. This almost always results in a series of hierarchically inclusive circles of increasing size but decreasing intimacy, as depicted in Fig.~\ref{Ego_Network_Model}. Both the number of circles and the sizes of the circles tend to be relatively unvarying, with the average sizes usually being found at around 5, 15, 50 and 150 Alters~\cite{Zhou_2005}. Similarly, the scale ratio between them is remarkably consistent, almost always being very close to 3~\cite{Hill_2003}.

The importance of the ENM comes in large part from how omnipresent it is in social networks. Indeed, it has been found in a diverse range of social systems, including traditional hunter-gatherer groups and small-scale horticultural communities from across 5 continents as well as the armies of ancient Rome and most modern-day militaries~\cite{Dunbar_1993}. In fact, the ENM is so ubiquitous that it can even be observed in many species of non-human primates, albeit with smaller group sizes~\cite{Dunbar_1998}. The pervasiveness of the ENM is explained by Dunbar's Social Brain Hypothesis~\cite{Dunbar_1998}, which posits that primates have a hardwired cognitive limit that determines the maximum size and complexity of the social groups they can maintain; for humans, this is around 150 (Dunbar's number). Once this limit is surpassed, a social group will start to become unstable and fragment into smaller, more manageable groups~\cite{Dunbar_1992}. One might expect that the relative ease of communicating online would allow humans to maintain larger social networks. However, this doesn't appear to be the case and the ENM structure is very similar in online contexts, the only real difference being that an additional, innermost circle is sometimes found for OSNs with a mean size of around 1.5 Alters~\cite{Dunbar_2015} (the existence of which has been postulated for offline networks as well but has not been proven due to lack of data).
Because of how ubiquitous the ENM is, observing a network from the singular perspectives of Egos can reveal insights that are only visible at a microscopic scale, yet have far-reaching consequences across the entire network. Indeed, the structural properties of the ENM have been shown to influence a number of social behaviours, such as collaboration and information diffusion~\cite{Sutcliffe_2012}. 

Although the ENM can provide many additional insights, it still has some notable limitations. One such limit comes from how tie strength is measured. Traditionally, the strength of an Ego-Alter connection has been measured using the frequency of their interactions. While this has been shown to be a good proxy for the strength of a relationship~\cite{Gilbert_2009}, not all relationships can be differentiated merely by their strength. 
For instance, an individual with a supportive coworker and an angry neighbour will have two very different relationships. Indeed, even though the interaction frequencies, and therefore the tie strength, may even be very similar, the former relationship will be far more positive than the latter. 
One way of better accounting for the dualistic aspect of social networks is to use a signed representation, known as a \emph{signed network}. In a signed network, edges have a polarity (+/-): positive links indicate friendship, trust and similarity, while negative links are associated with hatred and distrust.  It has been shown that positive and negative relationships play very different roles within a network, which can be leveraged to improve many network-related tasks, such as community detection~\cite{Esmailian_2015} and opinion dynamics~\cite{Shi_2016}. In particular, negative links are significantly more informative than positive ones~\cite{Leskovec_2010_b_Predicting}.
Thus, the inclusion of signs may greatly enhance our understanding of the ENM and social networks in general. However, the sign of an OSN relationship is an implicit piece of information, which typically needs to be estimated by interaction data. Indeed, the sign prediction problem in OSNs is a research challenge per se.

This paper proposes two main contributions. The first contribution is a \emph{novel method, grounded in quantitative results from psychology, of inferring signed relationships in unsigned network data}, allowing an unsigned network to be converted into a signed one. This method (i) requires only text-baseda interactions to sign a relationship (hence, it can be applied to any network in which users interact principally via text, i.e. in the vast majority of popular OSNs), (ii) is designed for the short texts typical of OSNs interactions, (iii) requires only data about the interactions over the links we want to sign (hence scales linearly with them).

The second contribution is a \emph{thorough analysis of Signed Ego Networks (SEN), i.e. Ego Networks where edges have a polarity}. SENs are obtained for 8 Twitter datasets by first generating unsigned Ego Networks and then using the aforementioned method to apply signs to them. Again, the unsigned and signed versions are analysed, including the distribution of signed links across the various circles of the SEN.
The main findings are that: (i) Twitter users engage in much more negative relationships than expected in the Active Networks (illustrated in Fig.~\ref{Ego_Network_Model}); (ii) specialised users (e.g. journalists) do so to an even higher extent, and (iii) negative relationships are particularly present in the intimate EN layers of specialised users. All in all, the results confirm the popular notion that higher engagement in online social interactions results in being exposed to increasingly negative relationships and sentiments, which are propagated across the entire network.

\section{Background}
\subsection{Ego Network Model}
\label{sec:egonets_background}
As mentioned in the previous section, the ENM is a structure with an Ego at the centre, surrounded by groups of Alters in concentric circles. The ENM stems from the Social Brain Hypothesis~\cite{Dunbar_1998} in anthropology, which posits that the social life of primates is constrained by the size of their neocortex. Accordingly, the typical group size for humans is estimated to be around 150 individuals (the famous Dunbar's number). Note that these 150 ``friends'' a person engages with do not include acquaintances but only relationships that are regularly nurtured. In modern societies, this implies a minimum of exchanging birthday or holiday wishes once a year. These quality relationships constitute the so-called \emph{active} part of the Ego Network.

Unsurprisingly, the 150 relationships are not all equivalent and the Ego interacts with them at different frequencies. Specifically, Alters in the active network can be grouped into concentric circles~\cite{Hill_2003,Zhou_2005} based on their tie strength to the Ego. As mentioned in the previous section, both the number of circles (approximately 4 or 5) and their sizes -- $1.5, 5, 15, 50, 150$ -- are fairly regular, in offline and online social networks~\cite{Dunbar_2015}. The circles are conventionally concentric, with the inner circles being contained in the outer ones. 
Note also that, by definition, the size of the active part of the Ego Network is equivalent to the size of its outermost circle.

Since the Ego-Alter tie strength determines which Alters belong to which circle, the tie strength is a key concept in this model. It was defined by Granovetter as the equally weighted combination of 4 elements of a relationship: the time spent maintaining it, its emotional intensity, its level of intimacy and the reciprocal services it generates~\cite{Granovetter_1973}. This definition can be an important consideration when understanding the behaviours of various users. For instance, users that engage in OSNs for professional reasons may spend more time on social platforms thereby generating more reciprocal services and spending more time maintaining relationships. Indeed, it has previously been hypothesised that journalists are likely to be more cognitively involved with Twitter than other types of users~\cite{Toprak_2021_b_Region-based}. While being just one of the tie strength dimensions described by Granovetter, the time spent maintaining a relationship has been the exclusive focus of the related literature on Ego Networks, due to its widespread availability and ease of computation (using the number of interactions as its proxy). The goal of this work is to advance the state of the art by investigating the emotional intensity of relationships alongside the time spent maintaining them.

\subsection{Signed Networks}
\label{sec:signed_networks_background}
Unlike unsigned networks, whose connections between users are either binary (i.e., link existing/missing) or weighted (typically based on the tie strength), signed networks have connections that can be further distinguished as either positive or negative (sometimes called \emph{polarity} of edges~\cite{Tang_2016}). Along with the inherent assumption that positive links represent positive relationships, they have also been used to infer trust and homogeneity~\cite{Maniu_2011}. Similarly, negative links are used to infer negative relationships, distrust and dissimilarities. Thus, signed networks contain additional information that can be leveraged to improve the performance of many tasks, including community detection~\cite{Traag_2009} and information diffusion~\cite{Ferrara_2015}.
Furthermore, signed networks are known to follow certain properties, such as Balance Theory~\cite{Heider_1946}, which postulates that certain configurations of signed triads (i.e. groups of three users who are all interconnected by signed edges) should be more common than others when observed across an entire network.

Previous observations of networks with publicly available signed connections have found that negative connections tend to be much rarer than positive connections, making up around 15.0\% to 22.6\% of the total connections in a network~\cite{Leskovec_2010_a_Signed}. The fact that link polarity is known to the users in these networks might amplify social pressure and effects such as social capital~\cite{Coleman_1988}, whereby relationships between individuals who have many relationships in common are more likely to be positive due to social pressure from the surrounding community. In unsigned networks, where negative connections are not explicitly visible, this social pressure should be lower. Hence, it would be reasonable to expect that networks without explicitly signed relationships would have higher proportions of negative relations than what is observed in explicitly signed ones. We will test this hypothesis in section~\ref{sec:results}.

Despite the added advantages of signed networks, they are rarely the focus of research as the vast majority of popular social networks do not allow users to create explicitly negative links, meaning that signed network data are seldom readily available. Notable exceptions are Slashdot and Epinions (the latter now defunct), which have provided two of the most popular benchmark datasets for signed networks~\cite{Tang_2016}. Unfortunately, as they do not feature interaction frequencies, they cannot be leveraged for Ego Network analysis. ENM studies typically use Twitter data (due to their public nature and easy access via the Twitter API) but signs are not available for relationships on Twitter.
However, just as with real-world relationships, the online interactions of individuals usually contain implicit information about whether their connections are positive or negative~\cite{Maniu_2011}. % removed: Kunegis_2013 

Some methods have already been developed to predict the signs of unsigned networks. However, the majority of these have focused on using the structural elements of the surrounding network to determine the signs of relationships, for example, clustering coefficient~\cite{Javari_2014}. Similarly, methods have been established for signing a novel network using classification algorithms trained on previous datasets with known signs~\cite{Tang_2016}. % removed: Ye_2013, Yuan_2019
All these techniques have viewed the problem of sign prediction from a top-down perspective, looking at the features of a network as a whole and inferring signs based on the structure of the links. However, by taking the inverse approach, and viewing the problem from the bottom-up, it is possible to consider the more qualitative aspects of connections that have largely gone unexplored. 

The basic blocks that constitute a relationship are the interactions and exchanges between users and their sentiment. Sentiment analysis for individual exchanges is extremely well established~\cite{Liu_2012} and it is possible to apply signs to these bottom-level interactions with a high degree of confidence. However, ways of extending these singular labels to whole series of interactions, or relationships, are largely undeveloped in OSN research. One previous study that has examined this problem~\cite{Hassan_2012} employed a Support Vector Machine (SVM) to sign relationships between users in discussion forums based on 4 user features and 3 interaction features. The SVM was trained on an annotated dataset and is shown to achieve an accuracy above 0.835 on the test set. Unfortunately, this approach is not directly and reliably replicable for Twitter interactions, which are characterised by short textual interactions (much shorter than in discussion forums) and less structured. In addition, the ground truth dataset used for the training phase is not publicly available. Thus, here we propose an alternative approach that targets short texts and leverages a pretrained model from previous literature (VADER~\cite{Hutto_2014}) that is specifically designed for sentiment analysis on short texts. 

\subsection{Positive and negative relationships in psychology}
\label{sec:signed_relations_psychology}
It is perhaps unsurprising that psychology has also observed social interactions through the lens of positive and negative interactions. Indeed, many experimental and behavioural analyses have investigated the effects of positive and negative social exchanges in various contexts. One subarea of this research that is particularly pertinent investigates the effect of the ratios of positive and negative interactions. Indeed, an optimum ratio of negative interactions has been found for numerous relationship types: marriages that display levels of negativity beyond this ratio are less likely to last~\cite{Gottman_1995} and parents who communicate with more negativity are more likely to cause detrimental effects on their child's development~\cite{Hart_1995}. This ratio is usually observed at around 1 negative interaction for every 5 positive interactions, or roughly 17\% negative. We refer to this ratio as the \emph{golden interaction threshold} and we will leverage it in our proposed method for signing relationships.

\section{Methodology and Data}
\label{sec:methodology}
In this section we describe how to obtain Signed Ego Networks, assuming that Twitter data are used as input (this is the de-facto standard in the related ENM literature~\cite{Arnaboldi_2017,Dunbar_2015,Toprak_2021_a_Harnessing,Toprak_2021_b_Region-based}). For building Ego Networks, Tweets that involve a form of direct communication between Twitter users are needed. These are Replies, Mentions and Retweets. In them, users explicitly reply to other users' Tweets, mention other users (via the ``@'' symbol) in their Tweets, or retweet their content (i.e., share other users' Tweets on their own timeline, sometimes with a personal comment on the shared Tweets, which is known as a Quote Retweet). Each of these directed Tweets corresponds to an interaction between the Ego and an Alter. While some of these (re-)Tweets may be addressed to the wider network, beyond the specific Alter, they still denote a cognitive involvement of the Ego with the Alter, which is the most important feature to associate an interaction to a specific social relationship~\cite{Dunbar_1998}.

\subsection{Signing Relationships}
\label{sec:signing_relationships}
The method for assigning positive and negative signs to the relationships consists of two main steps. First, sentiment analysis is performed on each interaction between an Ego and Alter. Next, a sign is determined for the whole relationship based on the proportion of sentimentally negative interactions generated by the relationship.

\emph{Step 1: labelling single interactions--} The sentiment analysis of the individual interactions is performed using the well-established Valence Aware Dictionary and sEntiment Reasoner (VADER) library~\cite{Hutto_2014}. VADER is a sentiment analysis tool developed specifically for use with social media data. The creators of VADER found that it had the best precision and accuracy when compared to 7 state-of-practice alternatives, and also performed better than individual humans when analysing social media text~\cite{Hutto_2014}. VADER is considered one of the standard techniques for sentiment analysis of short text and is thus appropriate as a tool for the rest of our analysis.

We use VADER to label the Tweets that compose the text-based interactions between each pair of users (i.e., Replies, Mentions, and Quote Retweets). We choose to assign regular Retweets a neutral sentiment. This was done because Retweets were not originally written by the Egos, meaning that their sentiments are not indicative of the cognitive effort being invested by the Egos in their relationships with the Alters. Rather, Retweets can be better understood as an Ego's desire to share the content of an Alter. With regards to Granovetter's definition, Retweets can therefore be viewed as a reciprocal service generated by the relationship. Furthermore, treating them as sentimentally neutral interactions reduces their relative weighting towards the sign of the overall relationship, which reflects the lower cognitive and time costs of clicking the Retweet button compared to writing out a mention or a reply. However, if the individual added some additional text while retweeting (Quote Retweet), then the sentiment of this text is considered as it had been produced by the Ego. 

\emph{Step 2: labelling a relationship--} We assign labels to each relationship based on the proportion of negative interactions it has generated. Specifically, we use the golden interaction ratio (17\%), discussed in section~\ref{sec:signed_relations_psychology}, as a threshold. Relationships with a higher percentage of negative interactions are labelled as negative and those with equal or lower percentages are labelled as positive. 

\subsection{Computation of Signed Ego Networks}
\label{sec:computing_signed_egonets}
Computing the Ego Networks is done by calculating the frequency of interaction between each Ego-Alter pair and then clustering the Alters based on these frequencies. This is a well-established method and has previously been conducted using a variety of clustering algorithms: k-means~\cite{MacQueen_1967}, DBSCAN~\cite{Ester_1996}, MeanShift~\cite{Fukunaga_1975}. The latter approach is used in the current study as it is one of the more common options, and also automatically determines the best number of clusters. 

The signs of each relationship are computed separately, as described in the previous subsection (section~\ref{sec:signing_relationships}), and applied to the Ego Networks, after they have been computed. This allows us to investigate how the cognitive effort of Egos is allocated to positive and negative relationships across the social circles.

\subsection{Datasets}
\label{sub:datasets}
% General + Twitter
For the present study, a collection of 8 datasets was used. These datasets were all collected from Twitter, using the official Twitter API. They comprise the last 3,200 Tweets in the timeline of the selected Twitter users, i.e. the maximum allowed by the standard public API v1.1. This amount of Tweets has shown to be sufficient to generated sensible EN of users (e.g.~\cite{Arnaboldi_2017,Arnaboldi_2015,Dunbar_2015}). Of these Tweets, we keep the Replies, Mentions, and Retweets. 
These directed Tweets will be used to compute the frequency of interactions between Egos and Alters and to build the Signed Ego Networks as described in section~\ref{sec:methodology}. By only considering the tweets created by the Ego (and not those of the Alters), the data better reflect the cognitive and time constraints of the Ego.

% Generic and specialised users
These eight datasets represent a mix of specialised and generic users, the former being users who use Twitter for professional reasons and the latter being users who either use Twitter for personal reasons or are not explicitly known to use it for professional reasons. It is important to distinguish between these two types of users as several differences have previously been observed in the behaviour of certain specialised users (namely journalists~\cite{Toprak_2021_b_Region-based}) in online social contexts. Detailed descriptions of these datasets are available in the following subsections and the exact numbers of Egos, Alters, relationships and interactions can be seen in Table~\ref{descriptive_full} and Table~\ref{descriptive_active}, the former containing all collected users and the latter containing only the users that remained after the preprocessing steps detailed in section ~\ref{subsub:preprocessing}.

\begin{table*}[htbp]
\centering
\caption{Number of Egos, Alters, relationships and interactions in the full Ego Networks, before removing unengaged users (as described in section~\ref{subsub:preprocessing})}
\label{descriptive_full}
\begin{tabular}{|c|c|c|c|c|}
\hline
\textbf{Dataset} & \textbf{  Egos  } & \textbf{  Alters  } & \textbf{ Relationships } & \textbf{ Interactions }\\
\hline
American Journalists & 1,714 & 505,023 & 1,479,764 & 4,677,736\\
Australian Journalists & 957 & 185,245 & 709,764 & 2,466,111\\
British Journalists & 512 & 209,402 & 469,863 & 1,397,996\\
NYT Journalists & 678 & 173,620 & 521,917 & 1,493,199\\
Science Writers & 497 & 182,240 & 463,624 & 1,350,799\\
British MPs & 584 & 157,053 & 343,366 & 1,277,010\\
\hdashline
Monday Motivation & 6,946 & 1,151,899 & 2,291,692 & 9,449,775\\
UK Users & 3,512 & 12,088,975 & 2,507,634 & 9,931,908\\
\hline
\end{tabular}
\end{table*}

\begin{table*}[htbp]
\centering
\caption{Number of Egos, Alters, relationships and interactions in the active networks of each dataset, after removing unengaged users (as described in section~\ref{subsub:preprocessing})} 
\label{descriptive_active}
\begin{tabular}{|c|c|c|c|c|}
\hline
\textbf{Dataset} & \textbf{  Egos  } & \textbf{  Alters  } & \textbf{ Relationships } & \textbf{ Interactions }\\
\hline
American Journalists & 1,037 & 68,792 & 143,390 & 1,639,623\\
Australian Journalists & 520 & 26,561 & 75,455 & 937,764\\
British Journalists & 281 & 24,614 & 41,524 & 434,477\\
NYT Journalists & 558 & 23,327 & 59,922 & 561,563\\
Science Writers & 241 & 18,531 & 35,185 & 381,340\\
British MPs & 440 & 27,538 & 76,857 & 323,765\\
\hdashline
Monday Motivation & 1,461 & 78,906 & 158,374 & 894,648\\
UK Users & 921 & 84,993 & 111,426 & 1,474,882\\
\hline
\end{tabular}
\end{table*}

\subsubsection{Specialised Users}
\label{subsub:datasets_specialised}
\hfill \\
\emph{Journalists--}
The first 3 datasets had already been collected as part of a previous study that investigated the Ego Networks of journalists~\cite{Toprak_2021_b_Region-based}. The aforementioned study originally included 17 different datasets from all across the globe. However, in order to apply sentiment analysis to these datasets, we only use those with text data in English, those being journalists from the United States, Australia and the United Kingdom. The British dataset was collected on 31st January 2018 and the others were collected between the 14th and 15th of May 2018, using existing lists of journalists (validated in~\cite{Boldrini_2018}).

In addition to these, another set of journalist data were taken from a different study~\cite{Ollivier_2022}. This dataset was collected from a list of New York Times journalists, created by the New York Times itself. All the users from this list were downloaded on February 16th 2018. This dataset will be referred to as NYT Journalists.

\emph{Science Writers--}
This dataset was collected using a list created by a science writer at Scientific American, Jennifer Frazer. It was downloaded on June 20th 2018, as part of a previous study~\cite{Ollivier_2022}.

\emph{British Members of Parliament (MPs)--}
The British MPs dataset is a novel one that was collected as part of the present study. It includes the Timelines of members of the British Parliament, taken from a publicly available list provided by UKinbound~\cite{UK_inbound_2020}. These Timelines were collected between the 4th and 6th March 2022. The Tweet IDs for this dataset are provided at \url{https://zenodo.org/record/6420845}.

\subsubsection{Generic Users}
\label{subsub:datasets_generic}
\hfill \\
\emph{Monday Motivation--}
The first generic dataset consisted of the Timelines of users who tweeted or retweeted, in English, using the hashtag \#MondayMotivation on 16th January 2020. The Timelines were then collected between the 17th and 18th of January 2020, as part of a previous study~\cite{Ollivier_2022}. Bots and similar types of non-human users who are not constrained by the same cognitive and time constraints as regular humans were removed by the original authors.

\emph{UK Users--}
The second generic dataset came from a random sample of all users who tweeted or retweeted from the United Kingdom, in English, on February 11th 2020. The Timelines of these users were collected on 13th February 2020, as part of a previous study~\cite{Ollivier_2022}. As with the previous dataset, bot accounts were removed from this dataset by the original authors.

\subsubsection{Preprocessing} 
\label{subsub:preprocessing}
\hfill \\
% Inactive/irregular filters
Before conducting any analyses on the ENMs, it was necessary to filter out inactive and irregular users, who were unlikely to be engaged enough with Twitter to have fully developed Ego Networks on the platform. For this, Egos were removed if their timeline consisted of less than 2,000 tweets total, spanned a period of fewer than 6 months (from the first to the last tweet in their Timeline) or they tweeted less than once every 3 days for more than 50\% of the months that they were active. These filtration parameters are in line with those of previous work on Ego Networks~\cite{Arnaboldi_2015,Toprak_2021_b_Region-based}, to which we refer for further details.

\section{Results}
\label{sec:results}
In this section, we investigate the properties of the Signed Ego Networks of the 8 selected datasets extracted according to the methodology discussed in section~\ref{sec:methodology}. Recalling from section~\ref{sec:egonets_background} that an Ego Network is composed of an active and inactive part, we first study how negative relationships are distributed in the full vs active-only network in section~\ref{sec:results_full_active}. Then, in section~\ref{sec:results_specialised_generic}, we discuss the differences between specialised and generic users. Finally, in section~\ref{sec:results_circles}, we analyse how positive and negative relationships are distributed across the Ego Network social circles.

\subsection{Negative Relationships in Full and Active Networks}
\label{sec:results_full_active}
An initial comparison is carried out between the percentages of negative relationships in the full networks of each dataset and those of the active networks. As mentioned previously, the active network consists of the Alters with whom the Ego engages meaningfully, i.e., at least once a year according to anthropology results. The results of this comparison are summarised in Table~\ref{descriptive_negative}.

\begin{table*}[htbp]
\caption{Mean percentages of negative relationships in the full and active networks. A dashed line is used to separate specialised users (above) and generic users (below).}
\label{descriptive_negative}
\begin{center}
\begin{tabular}{|c|c|c|c|}
\hline
\textbf{Dataset} & \textbf{Full Negatives (\%)}$^{\mathrm{a}}$ & \textbf{Active Negatives (\%)$^{\mathrm{a}}$} & \textbf{Difference}\\
\hline
American Journalists & 27.15 [27.06, 27.23] & 47.97 [47.44, 47.96] & +20.82\\
Australian Journalists & 31.54 [31.41, 31.67] & 54.39 [53.91, 54.61] & +22.85\\
British Journalists & 28.78 [28.62, 28.94] & 50.37 [49.72, 50.68] & +21.59\\
NYT Journalists & 31.58 [31.43, 31.74] & 54.89 [54.49, 55.29] & +23.31\\
Science Writers & 25.62 [25.45, 25.78] & 45.23 [44.71, 45.75] & +19.62\\
British MPs & 19.24 [19.09, 19.38] & 29.03 [28.66, 29.39] & +9.79\\
\hdashline
Monday Motivation & 16.45 [16.36, 16.54] & 21.83 [21.58, 22.08] & +5.38\\
UK Users & 24.22 [24.13, 24.31] & 35.32 [35.04, 35.60] & +11.10\\
\hline
\multicolumn{4}{l}{$^{\mathrm{a}}$95\% confidence intervals shown in square brackets.}
\end{tabular}
\end{center}
\end{table*}

For the full networks, most of the datasets display quantities of negative relationships within or slightly above the expected 15.0\% to 22.6\% range~\cite{Leskovec_2010_a_Signed} previously mentioned in section~\ref{sec:signed_networks_background}, which supports the validity of the chosen datasets. Indeed, the full network with the highest percentages of negative relationships, NYT Journalists, was less than a third negative. In stark contrast, all of the active-only networks were significantly higher than the corresponding full networks, with some even surpassing 50\% negativity. This increase in negative relationships is more prominent for specialised users than for generic users.

The higher negative percentages of the active networks compared to the full networks suggests that Egos tend to have proportionally more negative relationships with Alters they engage with frequently than with acquaintances. It has previously been found that negative emotions and communications elicit a stronger response than positive ones~\cite{Baumeister_2001} and that negative entities appear to be more contagious~\cite{Rozin_2001}. Thus, one explanation for why the active networks are more negative is that, because the active users are spending more time engaged in online platforms, they have an elevated risk of being exposed to negativity, and therefore more likely to also spread negativity. This is further supported by the observation that the journalist datasets are the most negative as it has previously been hypothesised that journalists are likely to be more cognitively involved with Twitter than other types of users~\cite{Toprak_2021_b_Region-based}, such as politicians and generic users. Therefore, the effects of exposure to negativity may affect them more strongly, thus resulting in a bigger relative increase in the negativity of their active networks. This would also explain why the British MPs dataset had the lowest rate of negative relationships out of all the specialised networks.

In addition, given that the active-only networks had significantly higher rates of negative relationships across all 8 datasets, it appears that this increase in negative relationships for the active part of Ego Networks is not unique to any specific community, but rather a byproduct of communicating and engaging with Twitter. It may therefore be interesting for future research to investigate whether this effect is similar for other OSNs and how the effects compare across different social platforms.

\subsection{Negative Relationships of Specialised and Generic Users}
\label{sec:results_specialised_generic}

Next, the percentages of negative relationships between the networks of specialised and generic users were compared. As can be seen in Table~\ref{descriptive_negative}, most of the specialised users displayed more negative relationships than the generic users. However, this difference was fairly small for the full networks, with the generic UK Users dataset (24.22\%) actually containing more negative relationships than the British MPs (19.24\%) and nearly as much as the Science Writers (25.62\%).
By contrast, the active networks of the specialised users were much more negative, with the only exception to this being the British MPs dataset, whose change in negativity better matched those of the generic datasets. Indeed, barring the British MPs, the next least negative specialised dataset, Science Writers (45.23\%), was nearly 10 percentage points more negative than the most negative generic dataset, UK Users (35.32\%). This is particularly surprising given that specialised users are more likely to regularly work or interact in close proximity and therefore should be subject to stronger social pressures to maintain (at least the appearance of) positive relationships, an effect known as social capital \cite{Coleman_1988}.

\subsection{Circle-by-Circle Analysis of the Signed ENM}
\label{sec:results_circles}

After calculating the Ego Networks of each dataset, the circle sizes and negative relationship percentages were able to be compared at each level of the ENMs. As mentioned previously, the circles of the ENM are concentric, meaning that each one contains all the Alters of those preceding it.
It is important to note that different Egos may have slight variances in their number of optimum Dunbar circles, as well as the size of their Ego Network, due to individual differences. Similar differences have been observed repeatedly in previous research~\cite{Arnaboldi_2013,Dunbar_2015}, therefore, in order to standardise the results between users when analysing the different Dunbar circles, it is common practice to focus principally on Egos who have a common optimum circle number~\cite{Toprak_2021_a_Harnessing,Toprak_2021_b_Region-based}. As can be seen in Table~\ref{optimum_circles_and_mean_ego_sizes}, the closest whole number to each of the mean optimum circle numbers is 5 for all except two of the datasets, the exceptions being NYT Journalists and British MPs. Five has also previously been found to be a common optimum number of circles for Twitter users~\cite{Arnaboldi_2017,Dunbar_2015}. Therefore, only Ego's whose optimum number of circles was 5 were considered for the subsequent circle-by-circle analyses\footnote{This explains why the active Ego Network sizes in Table~\ref{optimum_circles_and_mean_ego_sizes} do not match the Circle 5 sizes in Table~\ref{circle_all}: the latter is computed on a subset of users, those with optimum circle number equal to 5.}.

\begin{table*}[htbp]
\centering
\caption{Mean Ego Network sizes and mean optimum number of circles in the active networks for each dataset} \label{optimum_circles_and_mean_ego_sizes}
\begin{tabular}{|c|c|c|}
\hline
\textbf{Dataset} & \textbf{Mean Optimum Circle}$^{\mathrm{a}}$ & \textbf{Mean Ego Network Size}$^{\mathrm{a}}$\\
\hline
American Journalists & 5.30 [5.22, 5.38] & 138.27 [134.26, 142.29]\\
Australian Journalists & 5.17 [5.06, 5.28] & 145.11 [139.55, 150.66]\\
British Journalists & 5.42 [5.27, 5.56] & 147.77 [140.47, 155.08]\\
NYT Journalists & 5.53 [5.40, 5.66] & 155.67 [148.61, 162.74]\\
Science Writers & 5.44 [5.29, 5.60] & 148.45 [148.45, 156.77]\\
British MPs & 6.00 [5.87, 6.12] & 174.68 [168.13, 181.22]\\
\hdashline
Monday Motivation & 5.07 [5.00, 5.15] & 107.42 [103.83, 111.00]\\
UK Users & 5.23 [5.14, 5.32] & 120.98 [115.75, 126.22]\\
\hline
\multicolumn{3}{l}{$^{\mathrm{a}}$95\% confidence intervals shown in square brackets.}
\end{tabular}
\end{table*}

First of all, the mean sizes of each circle were examined. As expected, the sizes are close to those of Dunbar's expected values: i.e. 1.5, 5, 15, 50, 150~\cite{Dunbar_2015}, the exact numbers can be seen in Table~\ref{circle_all}. A few of the datasets become more distant from the expected numbers in the outer circles, however, this has also been observed in previous research~\cite{Arnaboldi_2017,Toprak_2021_b_Region-based} and can be attributed to even the most engaged users not developing the entirety of their social network online.

\begin{table*}[htbp]
\centering
\caption{Mean circle sizes of Egos with an optimum circle number of 5}
\label{circle_all}
\begin{tabular}{|c|c|c|c|c|c|}
\hline
\textbf{Dataset} & \textbf{Circle 1} & \textbf{Circle 2} & \textbf{Circle 3} & \textbf{Circle 4} & \textbf{Circle 5}\\
\hline
American Journalists & 1.61 & 5.33 & 15.01 & 41.78 & 127.28\\
Australian Journalists & 1.41 & 4.76 & 13.61 & 40.22 & 134.71\\
British Journalists & 1.83 & 6.27 & 16.87 & 48.07 & 142.52\\
NYT Journalists & 1.65 & 5.43 & 14.76 & 40.16 & 114.68\\
Science Writers & 1.70 & 5.81 & 16.40 & 44.29 & 124.86\\
British MPs & 1.98 & 6.67 & 18.09 & 49.00 & 146.79\\
\hdashline
Monday Motivation & 1.72 & 5.26 & 13.22 & 33.58 & 103.71\\
UK Users & 1.84 & 5.96 & 15.72 & 39.32 & 114.66\\
\hline
\end{tabular}
\end{table*}

Next, the mean numbers and percentages of negative relationships were observed for each circle, these can be seen in Table~\ref{circle_negative}. It was found that the proportions of negative relationships were disproportionately higher at the more intimate levels of the ENM and decreased steadily towards the outer circles. All of the journalist datasets have negative percentages exceeding 60\% in the innermost circle and below 55\% in the outermost. This is very surprising as the inner circles are thought to hold more trustworthy and similar individuals (from the Ego's perspective). Indeed, one of the core components from Granovetter's definition of tie strength is reciprocal services~\cite{Granovetter_1973}, and reciprocity is thought to be very closely related to trust~\cite{Ostrom_2003}. The findings are even more surprising considering that the aforementioned effect of social capital would create a bias towards maintaining positive connections that would be strongest in the innermost circles, where the most interconnected users are expected to be.

\begin{table*}[htbp]
\centering
\caption{Number of negative relationships at each level of the Signed Ego Network (for Egos with an optimum circle number of 5)}
\label{circle_negative}
\begin{tabular}{|c|c|c|c|c|c|}
\hline
\textbf{Dataset} & \textbf{Circle 1}$^{\mathrm{a}}$ & \textbf{Circle 2}$^{\mathrm{a}}$ & \textbf{Circle 3}$^{\mathrm{a}}$ & \textbf{Circle 4}$^{\mathrm{a}}$ & \textbf{Circle 5}$^{\mathrm{a}}$\\
\hline
American Journalists & 0.99 (61.37\%) & 3.15 (59.13\%) & 8.53 (56.85\%) & 22.28 (53.33\%) & 60.23 (47.32\%)\\
Australian Journalists & 1.09 (77.30\%) & 3.34 (70.14\%) & 9.08 (66.74\%) & 25.27 (62.82\%) & 73.03 (54.21\%)\\
British Journalists & 1.16 (63.33\%) & 3.63 (57.98\%) & 9.76 (57.85\%) & 27.02 (56.22\%) & 70.94 (49.77\%)\\
NYT Journalists & 1.11 (67.21\%) & 3.73 (68.66\%) & 9.90 (67.05\%) & 24.73 (61.59\%) & 60.43 (52.70\%)\\
Science Writers & 0.82 (48.39\%) & 2.90 (49.91\%) & 7.97 (48.59\%) & 21.31 (48.11\%) & 55.87 (44.75\%)\\
British MPs & 0.58 (29.41\%) & 1.88 (28.24\%) & 5.08 (28.07\%) & 13.09 (26.71\%) & 31.31 (21.33\%)\\
\hdashline
Monday Motivation & 0.30 (17.72\%) & 0.97 (18.37\%) & 2.46 (18.59\%) & 5.79 (17.25\%) & 14.09 (13.58\%)\\
UK Users & 0.64 (34.75\%) & 2.00 (33.46\%) & 5.27 (33.54\%) & 13.14 (33.41\%) & 37.63 (32.81\%)\\
\hline
\multicolumn{5}{l}{$^{\mathrm{a}}$ Percentages in parentheses.}
\end{tabular}
\end{table*}

Finally, comparing the negative proportions between the user types, there appears to be a divide between specialised and generic users. Furthermore, this difference becomes even more noticeable when the journalists are compared to the non-journalists. The circle-to-circle variations appear to be much larger for the journalists than for the other types of users, with the most stable journalist dataset (British Journalists) dropping by 13.56 percentage points from Circle~1 to Circle 5. By contrast, the biggest variation for the non-journalists is 8.08 percentage points (British MPs).

Egos engage the most with their innermost circles, so any negativity coming from Alters in these intimate groups is more likely to be both seen and spread by the Ego, which will negatively affect the content and sentiment of the Ego's interactions. This also lends support to the notion that increased levels of engagement with Twitter lead to increased levels of negativity.

\section{Conclusion}
This paper has established a novel methodology for inferring signs for unsigned networks that can be employed whenever text-based communications between individual users are available. This methodology has a strong theoretical foundation and could allow future research to apply signed network techniques to non-signed networks. This methodology was used to generate signed relationships and Ego Networks for 8 datasets, which have been examined and compared to their unsigned counterparts. This has concluded in three main findings: (i) percentages of negative relationships tend to be higher for active-only networks than for full networks and this is more pronounced for specialised users than for generic users; (ii) specialised users display a higher propensity towards having negative relationships than generic users; (iii) negative relationships are found disproportionately more at the more intimate levels of the ENM.

There are several directions for future research, enabled by these findings. First, the results of this paper were consistent across numerous datasets, however, all the datasets were acquired from the same platform (Twitter). This raises the question of whether the observed effects are similar for other sources or whether they are unique to Twitter. Establishing an overall model of EN in general Online Social Networks is an exciting topic for future studies.

The definition of negative relationships in the literature broadly includes both ``positive relationships that share negative content'' and ``actually negative relationships'', and this is also the way we define them in the paper. It would be very interesting to understand to what extent sharing of negative content leads, possibly over time, to lack of trust and hatred even in positive relationships. This would require a large-scope, multidisciplinary study across social networks and human behavioural models.

Finally, it would be interesting to compare our sentiment-based method of signing a network to previously-established structural methods. Unfortunately, this would require a large, interconnected dataset that also contained the time and text of individual interactions between users, as well as having a ground truth about the polarity of the relationships (either from the users themselves or made by independent annotators). As far as we are aware, no such dataset is publicly available, and so such a comparison is currently impossible.

\section*{Acknowledgement}
This work was partially supported by the European Commission H2020 research and innovation programme, under grant agreements No 952026 (project HumanE AI Network) and No 871042 (project SoBigData++).

This work was also partially supported by the SAI project, under the CHIST-ERA grant CHIST-ERA-19-XAI-010, funded by MUR (grant No. not yet available), FWF (grant No. I 5205), EPSRC (grant No. EP/V055712/1), NCN (grant No. 2020/02/Y/ST6/00064), ETAg (grant No. SLTAT21096), BNSF (grant No. KP-06-DOO2/5).

\bibliographystyle{IEEEtran.bst}
\bibliography{biblio.bib}

\begin{thebibliography}{10}
\providecommand{\url}[1]{#1}
\csname url@samestyle\endcsname
\providecommand{\newblock}{\relax}
\providecommand{\bibinfo}[2]{#2}
\providecommand{\BIBentrySTDinterwordspacing}{\spaceskip=0pt\relax}
\providecommand{\BIBentryALTinterwordstretchfactor}{4}
\providecommand{\BIBentryALTinterwordspacing}{\spaceskip=\fontdimen2\font plus
\BIBentryALTinterwordstretchfactor\fontdimen3\font minus
  \fontdimen4\font\relax}
\providecommand{\BIBforeignlanguage}[2]{{%
\expandafter\ifx\csname l@#1\endcsname\relax
\typeout{** WARNING: IEEEtran.bst: No hyphenation pattern has been}%
\typeout{** loaded for the language `#1'. Using the pattern for}%
\typeout{** the default language instead.}%
\else
\language=\csname l@#1\endcsname
\fi
#2}}
\providecommand{\BIBdecl}{\relax}
\BIBdecl

\bibitem{Dunbar_1995}
R.~I. Dunbar and M.~Spoors, ``Social networks, support cliques, and kinship,''
  \emph{Human nature}, vol.~6, no.~3, pp. 273--290, 1995.

\bibitem{Zhou_2005}
W.-X. Zhou, D.~Sornette, R.~A. Hill, and R.~I. Dunbar, ``Discrete hierarchical
  organization of social group sizes,'' \emph{Proceedings of the Royal Society
  B: Biological Sciences}, vol. 272, no. 1561, pp. 439--444, 2005.

\bibitem{Hill_2003}
R.~A. Hill and R.~I. Dunbar, ``Social network size in humans,'' \emph{Human
  nature}, vol.~14, no.~1, pp. 53--72, 2003.

\bibitem{Dunbar_1993}
R.~I. Dunbar, ``Coevolution of neocortical size, group size and language in
  humans,'' \emph{Behavioral and brain sciences}, vol.~16, no.~4, pp. 681--694,
  1993.

\bibitem{Dunbar_1998}
R.~I.~M. Dunbar, ``The social brain hypothesis,'' \emph{Evolutionary
  Anthropology: Issues, News, and Reviews: Issues, News, and Reviews}, vol.~6,
  no.~5, pp. 178--190, 1998.

\bibitem{Dunbar_1992}
R.~I. Dunbar, ``Neocortex size as a constraint on group size in primates,''
  \emph{Journal of human evolution}, vol.~22, no.~6, pp. 469--493, 1992.

\bibitem{Dunbar_2015}
R.~I. Dunbar, V.~Arnaboldi, M.~Conti, and A.~Passarella, ``The structure of
  online social networks mirrors those in the offline world,'' \emph{Social
  networks}, vol.~43, pp. 39--47, 2015.

\bibitem{Sutcliffe_2012}
A.~Sutcliffe, R.~Dunbar, J.~Binder, and H.~Arrow, ``Relationships and the
  social brain: integrating psychological and evolutionary perspectives,''
  \emph{British journal of psychology}, vol. 103, no.~2, pp. 149--168, 2012.

\bibitem{Gilbert_2009}
E.~Gilbert and K.~Karahalios, ``Predicting tie strength with social media,'' in
  \emph{Proceedings of the CHI}, 2009, pp. 211--220.

\bibitem{Esmailian_2015}
P.~Esmailian and M.~Jalili, ``Community detection in signed networks: the role
  of negative ties in different scales,'' \emph{Scientific reports}, vol.~5,
  no.~1, pp. 1--17, 2015.

\bibitem{Shi_2016}
G.~Shi, A.~Proutiere, M.~Johansson, J.~S. Baras, and K.~H. Johansson, ``The
  evolution of beliefs over signed social networks,'' \emph{Operations
  Research}, vol.~64, no.~3, pp. 585--604, 2016.

\bibitem{Leskovec_2010_b_Predicting}
J.~Leskovec, D.~Huttenlocher, and J.~Kleinberg, ``Predicting positive and
  negative links in online social networks,'' in \emph{Proceedings of WWW},
  2010, pp. 641--650.

\bibitem{Granovetter_1973}
M.~S. Granovetter, ``The strength of weak ties,'' \emph{American journal of
  sociology}, vol.~78, no.~6, pp. 1360--1380, 1973.

\bibitem{Toprak_2021_b_Region-based}
M.~Toprak, C.~Boldrini, A.~Passarella, and M.~Conti, ``Structural models of
  human social interactions in online smart communities: the case of
  region-based journalists on twitter,'' \emph{arXiv preprint
  arXiv:2110.01925}, 2021.

\bibitem{Tang_2016}
J.~Tang, Y.~Chang, C.~Aggarwal, and H.~Liu, ``A survey of signed network mining
  in social media,'' \emph{ACM Computing Surveys (CSUR)}, vol.~49, no.~3, pp.
  1--37, 2016.

\bibitem{Maniu_2011}
S.~Maniu, T.~Abdessalem, and B.~Cautis, ``Casting a web of trust over
  wikipedia: an interaction-based approach,'' in \emph{Comp. proceedings of
  WWW}, 2011, pp. 87--88.

\bibitem{Traag_2009}
V.~A. Traag and J.~Bruggeman, ``{Community detection in networks with positive
  and negative links},'' \emph{Physical Review E}, vol.~80, no.~3, 2009.

\bibitem{Ferrara_2015}
E.~Ferrara and Z.~Yang, ``Quantifying the effect of sentiment on information
  diffusion in social media,'' \emph{PeerJ Computer Science}, vol.~1, p. e26,
  2015.

\bibitem{Heider_1946}
F.~Heider, ``Attitudes and cognitive organization,'' \emph{The Journal of
  psychology}, vol.~21, no.~1, pp. 107--112, 1946.

\bibitem{Leskovec_2010_a_Signed}
J.~Leskovec, D.~Huttenlocher, and J.~Kleinberg, ``Signed networks in social
  media,'' in \emph{Proceedings of the CHI}, 2010, pp. 1361--1370.

\bibitem{Coleman_1988}
J.~S. Coleman, ``Social capital in the creation of human capital,''
  \emph{American journal of sociology}, vol.~94, pp. S95--S120, 1988.

\bibitem{Javari_2014}
A.~Javari and M.~Jalili, ``Cluster-based collaborative filtering for sign
  prediction in social networks with positive and negative links,'' \emph{ACM
  TIST}, vol.~5, no.~2, pp. 1--19, 2014.

\bibitem{Liu_2012}
B.~Liu, ``Sentiment analysis and opinion mining,'' \emph{Synthesis lectures on
  human language technologies}, vol.~5, no.~1, pp. 1--167, 2012.

\bibitem{Hassan_2012}
A.~Hassan, A.~Abu-Jbara, and D.~Radev, ``Extracting signed social networks from
  text,'' in \emph{Workshop Proceedings of TextGraphs-7}, 2012, pp. 6--14.

\bibitem{Hutto_2014}
C.~Hutto and E.~Gilbert, ``Vader: A parsimonious rule-based model for sentiment
  analysis of social media text,'' in \emph{Proceedings of ICWSM}, vol.~8,
  2014, pp. 216--225.

\bibitem{Gottman_1995}
J.~Gottman, J.~M. Gottman, and N.~Silver, \emph{Why marriages succeed or fail:
  And how you can make yours last}.\hskip 1em plus 0.5em minus 0.4em\relax
  Simon and Schuster, 1995.

\bibitem{Hart_1995}
B.~Hart and T.~R. Risley, \emph{Meaningful differences in the everyday
  experience of young American children.}\hskip 1em plus 0.5em minus
  0.4em\relax Paul H Brookes Publishing, 1995.

\bibitem{Arnaboldi_2017}
V.~Arnaboldi, M.~Conti, A.~Passarella, and R.~I. Dunbar, ``Online social
  networks and information diffusion: The role of ego networks,'' \emph{Online
  Soc. Netw. Media}, vol.~1, pp. 44--55, 2017.

\bibitem{Toprak_2021_a_Harnessing}
M.~Toprak, C.~Boldrini, A.~Passarella, and M.~Conti, ``Harnessing the power of
  ego network layers for link prediction in online social networks,''
  \emph{IEEE Trans. Comput. Soc. Sys.}, 2022.

\bibitem{MacQueen_1967}
J.~MacQueen, ``Some methods for classification and analysis of multivariate
  observations,'' in \emph{Proceedings of the fifth Berkeley symposium on
  mathematical statistics and probability}, vol.~14.\hskip 1em plus 0.5em minus
  0.4em\relax Oakland, CA, USA, 1967, pp. 281--297.

\bibitem{Ester_1996}
M.~Ester, H.-P. Kriegel, J.~Sander, and X.~Xu, ``A density-based algorithm for
  discovering clusters in large spatial databases with noise,'' in \emph{kdd},
  vol.~96, 1996, pp. 226--231.

\bibitem{Fukunaga_1975}
K.~Fukunaga and L.~Hostetler, ``The estimation of the gradient of a density
  function, with applications in pattern recognition,'' \emph{IEEE Trans. on
  Inf. Theory}, vol.~21, no.~1, pp. 32--40, 1975.

\bibitem{Arnaboldi_2015}
V.~Arnaboldi, A.~Passarella, M.~Conti, and R.~I. Dunbar, \emph{Online social
  networks: human cognitive constraints in Facebook and Twitter personal
  graphs}.\hskip 1em plus 0.5em minus 0.4em\relax Elsevier, 2015.

\bibitem{Boldrini_2018}
C.~Boldrini, M.~Toprak, M.~Conti, and A.~Passarella, ``Twitter and the press:
  an ego-centred analysis,'' in \emph{Companion Proceedings of the The Web
  Conference 2018}, 2018, pp. 1471--1478.

\bibitem{Ollivier_2022}
K.~Ollivier, C.~Boldrini, A.~Passarella, and M.~Conti, ``Structural invariants
  and semantic fingerprints in the" ego network" of words,''
  \emph{arXiv:2203.00588}, 2022.

\bibitem{UK_inbound_2020}
``{List of MP Twitter Accounts},''
  \url{https://www.ukinbound.org/resources/list-of-mp-twitter-accounts/}, last
  accessed: 03 Mar 2022.

\bibitem{Baumeister_2001}
R.~F. Baumeister, E.~Bratslavsky, C.~Finkenauer, and K.~D. Vohs, ``Bad is
  stronger than good,'' \emph{Review of general psychology}, vol.~5, no.~4, pp.
  323--370, 2001.

\bibitem{Rozin_2001}
P.~Rozin and E.~B. Royzman, ``Negativity bias, negativity dominance, and
  contagion,'' \emph{Personality and social psychology review}, vol.~5, no.~4,
  pp. 296--320, 2001.

\bibitem{Arnaboldi_2013}
V.~Arnaboldi, M.~Conti, A.~Passarella, and F.~Pezzoni, ``Ego networks in
  twitter: an experimental analysis,'' in \emph{Proceedings IEEE INFOCOM},
  2013, pp. 3459--3464.

\bibitem{Ostrom_2003}
E.~Ostrom, ``Toward a behavioral theory linking trust, reciprocity, and
  reputation.'' \emph{Trust and reciprocity: Interdisciplinary lessons from
  experimental research}, pp. 19--79, 2003.

\end{thebibliography}

\end{document}